# Is a recently proposed experiment to demonstrate quantum behavior for optically levitated nanospheres feasible?

C. L. Herzenberg

**Abstract**
A recently proposed experiment considers the possibility of reaching regimes where quantum behavior might be observed in nano-mechanical systems. This proposed experiment is examined here for feasibility on the basis of results of earlier studies identifying a boundary separating obligatory classical behavior from quantum behavior based on effects dependent on large-scale properties of the universe. Calculations indicate that cosmologically-induced effects leading to a quantum to classical transition will not interfere with the proposed experiment at the level at which it is described. Thus, this experiment may be expected to be able to succeed for the case of nano-mechanical systems such as the 50 nanometer radius spheres under consideration; however, the success of similar experiments for larger micro scale systems may be ruled out.

**Introduction**

The presence of quantum behavior has been observed and demonstrated for a variety of small physical objects, including electrons, atoms, ions, and molecules, and including larger molecules such as fullerenes. A recent paper proposes setting up an experiment to reach a regime where quantum behavior might be observed in somewhat larger objects, in particular nano-mechanical systems.[1]

Over the past several years, evidence has accumulated that certain characteristics of our universe, notably its expansion and associated finite lifetime, can affect quantum objects in such a manner as to localize extended wave functions and their associated quantum objects into compact regions.[2-5] A number of studies have been examining aspects of the quantum behavior of objects that seem to be affected by characteristics of the universe such as the Hubble expansion and the finite duration of the universe since its inception at the Big Bang. These studies seem to substantiate that quantum objects and their wave functions may be limited in their spatial extent by cosmological effects. It follows from these studies that while smaller objects may retain quantum behavior, larger objects can be forced into exhibiting classical behavior.[2-5] Furthermore, the appearance of quantum phenomena can be restricted and changed into classical phenomena. The boundary between allowed quantum behavior and obligatory classical behavior based on these effects can be expressed in terms of a threshold moment of inertia.[4] By comparison of the magnitude of the moment of inertia of an object with the threshold moment of inertia, it should be possible to arrive at a determination of whether a particular object would be expected to be obligatorily classical due to these effects, or instead might exhibit quantum behavior under suitable conditions.

In the present article, we apply these considerations to evaluate the feasibility of the proposed new experiments to observe quantum behavior in nano-mechanical systems.

**Analysis and discussion**

A recent article, "Cavity optomechanics using an optically levitated nanosphere," by D. E. Chang, C. A. Regal, S. B. Papp, D. J. Wilson, J. Ye, O. J. Painter, H. J. Kimble, and P. Zoller, proposes experiments to create and demonstrate quantum behavior in optically levitated nanospheres.[1] We address the question as to whether these authors may in fact be able to demonstrate quantum behavior in nano-mechanical systems by such experiments, or whether the presence of what appears to be a fundamental constraint on quantum objects might prevent quantum behavior in such experiments.

Chang and colleagues remark that one of the most intriguing questions associated with quantum theory is whether effects such as quantum coherence and entanglement can be observed at mesoscopic or macroscopic scales. Chang and colleagues propose optically levitating a nano-mechanical system in an optical cavity to reach regimes where such quantum behavior might be observed. Through the long coherence times allowed, they suggest that this approach will enable ground-state cooling and coherent manipulation of a single mesoscopic mechanical system or entanglement generation between spatially separate systems. In particular, they explore the possibility of achieving these goals when the mechanical mode consists of the center-of-mass motion of an optically levitated dielectric nanosphere. Their dielectric sphere interacts with two standing-wave optical modes in a Fabry-Perot cavity; one resonantly driven mode provides an optical dipole trap for the nanosphere, while the second mode provides for radiation pressure cooling.[1]

Chang et al. examine their system with great care. They remark on how critical it is that the thermalization and decoherence rates of these systems be minimized. While they address these and many other concerns, the issue of whether recently examined cosmic effects that appear to be able to force quantum objects to classicality is not explicitly addressed, so we shall do so here.

Over the past several years, evidence has accumulated that certain characteristics of the expanding universe can affect quantum wave functions and the associated quantum objects in such a manner as to localize extended wave functions into compact regions, thus leading to classical behavior.[2-5] This appears to be a fundamental process.[4,5] These studies examining possible inherent limitations on quantum behavior for objects existing in an expanding universe of finite duration seem to have shown that sufficiently large objects must behave classically.[2-5]. These effects would be intrinsic, and could not be prevented or minimized by careful experiment design.

The threshold between classical and quantum behavior or quantum-classical boundary associated with these effects is sensitive to both size and mass, and turns out to depend very simply on what amounts to a combination of size and mass values in the form of a threshold moment of inertia. The threshold moment of inertia, which is associated with these cosmic effects that limit the appearance of quantum behavior and force the quantum object into classical behavior, has been estimated to be given approximately by the equation:[4]

$$I_{th} \approx h/4\pi H_o \qquad (1)$$

Here, h is Planck's constant and $H_o$ is the Hubble constant. While this expression can provide a useful threshold for separating obligatory classical behavior from possible quantum behavior of physical objects, it should be noted that fairly rough approximations were used in all of the studies addressing this topic, and as a result this resultant criterion itself is a rough estimate and can therefore be expected to provide a threshold value good to only an order of magnitude or so at best.

Eqn. (1) however does have the advantage of providing a very straightforward criterion for a seemingly important boundary separating the smaller objects that potentially exhibit quantum behavior from those larger objects that will necessarily exhibit classical behavior due to effects of the universe as a whole. Eqn. (1) identifies a threshold separating obligatory classical behavior imposed by cosmological effects on objects having large moments of inertia from possible quantum behavior for objects having smaller moments of inertia.

We can evaluate this threshold numerically; we will use h = 6.63 x $10^{-34}$ joule-seconds as the value for Planck's constant, and $H_o$ = 2.3 x $10^{-18}$ $sec^{-1}$ as a value for the Hubble constant. Inserting these values into Eqn. (1), we can obtain a numerical value for the threshold moment of inertia in mks or SI units as:

$$I_{th} \approx 2.3 \times 10^{-17} \text{ kg·m}^2 \qquad (2)$$

This result tells us that, approximately speaking, any object with a moment of inertia larger than about $10^{-17}$ kg·m$^2$ would be expected to behave in a classical manner, while any object with a moment of inertia smaller than about $10^{-17}$ kg·m$^2$ may exhibit quantum behavior, and would be expected to do so unless brought into classicality by other effects such as various quantum decoherence effects.[6-9]

We know that in general very small objects such as electrons, atoms, and small molecules behave quantum mechanically as entire objects. Small molecules have relatively small moments of inertia; for example, the moments of inertia for a water molecule with respect to different axes through the center of mass are reported to be in the range of about 1 x $10^{-47}$ kg·m$^2$ to 3 x $10^{-47}$ kg·m$^2$.[10] These moment of inertia values are some 30 orders of magnitude smaller than the critical threshold moment of inertia evaluated above, and thus are well within the range of expected quantum behavior according to this criterion.[4]

The largest objects for which successful quantum interference experiments have been reported are medium-sized molecules, the fullerenes.[8, 11-13] For orientation purposes, a $C_{60}$ fullerene buckyball has a diameter of about a nanometer. Quantum interference experiments with fullerenes (both $C_{60}$ and $C_{70}$ molecules) have been carried out.[8,11-13] Research groups have sent fullerene molecules with 60 or 70 carbon atoms each through the equivalent of two-slit interference equipment, dramatically displaying their quantum wave nature as entire objects in translational motion. These quantum interference experiments have established clearly that these intermediate size molecules can behave quantum mechanically with respect to their translational motion. A value for the moment of inertia of a fullerene buckyball ($C_{60}$) has been referred to in the literature as $1.0 \times 10^{-43}$ kg·m$^2$; additional measurements have been reported for other fullerenes.[13] These moment of inertia values for these medium-sized molecules that have been shown to exhibit superposed quantum states are roughly 26 orders of magnitude smaller than the quantum-classical boundary estimated above.

Since there is such a large range for values of moments of inertia above those of objects for which quantum interference has already been demonstrated but still below the threshold moment of inertia, it would seem that we can conclude that many physical structures considerably larger than fullerenes could also exhibit quantum interference effects, according to this criterion.

Chang and his colleagues suggest conducting quantum optomechanical experiments on objects levitated in vacuum inside an optical cavity, in particular by optically levitating a nano-mechanical system inside a Fabry-Perot optical cavity. They present evidence that the center-of-mass motion of a levitated nanosphere can be optically self-cooled to the ground state starting from room temperature. They analyse the possibility of performing the proposed experiment on nano-mechanical systems.[1]

In particular, Chang et al. consider an optically levitated dielectric sphere of radius 50 nm. Such nanospheres would be roughly a factor of 100 larger than fullerene buckyballs, the largest objects for which successful quantum interference experiments have so far been reported, as noted above. We can estimate the magnitude of the moment of inertia of such a nanosphere; it could be expected to be roughly in the range of $10^{-33}$ kg·m$^2$, depending on the density of the dielectric material of which the sphere is composed.

Thus, an object such as one of these nanospheres would have a moment of inertia some 10 orders of magnitude larger than that of a fullerene molecule, the largest type of object for which quantum interference experiments have so far been demonstrated.

However, judging by our present criterion given in Eqn. (2), such a nanosphere would be well below the threshold of obligatory classical behavior; - and below that threshold by roughly 16 orders of magnitude. Hence it would seem safe to conclude that these nanospheres would be well within the range of possible demonstrable quantum behavior unless brought into classicality by other effects.[4]

Of course, quantum objects can be brought into classicality by effects other than the limitations imposed by the universe that we have discussed here; however, cosmic effects cannot be obviated by careful experimental procedures that can, for example, minimize the decoherence brought about by thermal radiation.[8] Notably, various decoherence effects including those dependent on interactions with the local environment as well as other effects can bring about classical behavior in objects that would otherwise behave quantum mechanically.[6-9] It is interesting that Chang and colleagues in fact consider going beyond their initial proposed experiments so as to controllably study the decoherence of a large system.[1]

We close by emphasizing that a considerable range of larger nanoscale and mesoscale objects may be allowed to exhibit demonstrable quantum behavior in accordance with the criterion that we have discussed; however, as attention turns to possibilities for demonstrating quantum effects for larger systems in the range of micromechanical components and systems still below macroscopic sizes, the quantum-classical threshold associated with the cosmologically induced effects discussed here may be approached or exceeded. These effects should be taken into account in planning for experiments involving such larger systems, as has been pointed out recently in connection with proposed experiments considering the creation and demonstration of quantum superposition of small biological organisms.[14]

**Summary and conclusions**

The initial proposed experiment using dielectric nanospheres of 50 nanometer radius would be feasible rather than ruled out by quantum-classical transitions associated with cosmological effects dependent on the expansion or finite lifetime of the universe. Nano-mechanical systems of somewhat larger size may also exhibit quantum behavior in experiments, as may some systems in the mesoscopic size range. However, the quantum-classical boundary associated with cosmological expansion would appear to set a fundamental upper limit restricting the behavior of objects that are larger but still below macroscopic size.

quantum nanosphere size effects.doc
22 December 2009 draft